\documentclass[a4paper]{article}

\usepackage{INTERSPEECH2021}

\title{RW-Resnet: A Novel Speech Anti-Spoofing Model Using Raw Waveform}
\name{Youxuan Ma, Zongze Ren, Shugong Xu}
\address{
  Shanghai Institute for Advanced Communication and Data Science (SICS), Shanghai University,\\
  Shanghai, 200444, China}
\email{\{mayouxuan, zongzeren, shugong\}@shu.edu.cn}

\begin{document}

\maketitle
\begin{abstract}

In recent years, synthetic speech generated by advanced text-to-speech (TTS) and voice conversion (VC) systems has caused great harms to automatic speaker verification (ASV) systems, urging us to design a synthetic speech detection system to protect ASV systems. In this paper, we propose a new speech anti-spoofing model named ResWavegram-Resnet (RW-Resnet). The model contains two parts, Conv1D Resblocks and backbone Resnet34. The Conv1D Resblock is based on the Conv1D block with a residual connection. For the first part, we use the raw waveform as input and feed it to the stacked Conv1D Resblocks to get the ResWavegram. Compared with traditional methods, ResWavegram keeps all the information from the audio signal and has a stronger ability in extracting features. For the second part, the extracted features are fed to the backbone Resnet34 for the spoofed or bonafide decision. The ASVspoof2019 logical access (LA) corpus is used to evaluate our proposed RW-Resnet. Experimental results show that the RW-Resnet achieves better performance than other state-of-the-art anti-spoofing models, which illustrates its effectiveness in detecting synthetic speech attacks.


\end{abstract}

\section{Introduction}

Speaker voice can be used to confirm a person's identity, making automatic speaker verification (ASV) become a research hot-spot. But in real-world applications, ASV systems are at the risk of various spoofing attacks. The attacker can deceive ASV systems by taking advantage of the voice characteristics of a real speaker to disguise the identity of the target speaker \cite{wu2015spoofing}. 
Synthesis speech generated by advanced text-to-speech (TTS) \cite{oord2016wavenet, shen2018natural, prenger2019waveglow, wang2017tacotron} and voice conversion (VC) systems \cite{tanaka2019atts2s, zhang2019non} has become indistinguishable from the real speech of human in recent years, urging us to design an anti-spoofing model to protect the ASV system from synthesis speech attacks.


With the launch of ASVspoof Challenges \cite{wu2015asvspoof, Kinnunen2017, Todisco2019}, more and more researchers participate in spoofing attack countermeasure studies using different combinations of features and backbones \cite{Lavrentyeva2019, Alzantot2019, Lei2020, Tak2020, Wu2020}. 
The given log power spectrum (LPS) input is transformed to a genuinized feature in \cite{Wu2020} for detecting the spoofing attacks. \cite{tak2020end} delivers superior performance using the optimised linear frequency cepstral coefficients (LFCC). Constant Q cepstral coefficients (CQCC) and LFCC with the Siamese Convolutional Neural Networks (CNNs) are used in \cite{Tak2020} to improve the performance of CNN based anti-spoofing systems. 
However, some information is discarded in the process of extracting these hand-craft features. 
For example, the phase information is ignored when the model applies Fast Fourier Transform (FFT) to extract the feature of an audio signal.


Recently, directly using raw waveform as system input attracts more researchers. \cite{Jung2019} uses a modified CNN-GRU for the replay attack detection with the raw waveform as input.
The adapted Rawnet2 \cite{tak2020end} utilizes one-dimensional convolution layers and batch normalization (BN) for synthesis speech anti-spoofing. But these raw audio processing methods can not compare with traditional features while they are only using one-dimensional convolution layers.
On AudioSet tagging tasks, \cite{kong2020panns} proposes Wavegram. It is a neural network based time-frequency representation which is similar to Mel-spectrogram. And the corresponding models, Wavegram-CNN and Wavegram-Logmel-CNN, demonstrate strong competitiveness.


Inspired by all above, we argue that the potential of raw waveform has not been fully stimulated. It is a trend that using raw waveform as input, so the model can be trained end-to-end. Therefore, we explore new possibilities for speech anti-spoofing by using a novel neural network to extract features from raw waveform, avoiding the complicated feature extraction process. 
To sum up, the main contributions of our paper are as follows:

\begin{itemize}
\item We use Conv1D blocks to obtain the two-dimensional Wavegram feature from the raw waveform to replace the complicated process of hand-craft feature extraction. 
As we know, it is the first try of getting the neural network based time-frequency feature representation by directly using the raw waveform as input in the Speech Anti-Spoofing field.

\item On the basis of Conv1D block, we propose the Conv1D Resblock by adding a residual connection. The new path covers long-term features while the original path focuses on short-term features. 
By stacking the proposed Conv1D Resblocks, we get a better feature extractor for extracting features from the raw waveform. And the feature we get is called ResWavegram.

\item 
We propose a novel Anti-Spoofing model named ResWavegram-Resnet (RW-Resnet) and train it end-to-end.
Compared with other speech anti-spoofing models based on traditional hand-craft features, the proposed RW-Resnet makes a great improvement.
The result on the ASVspoof2019 logical access (LA) corpus shows that our model achieves state-of-the-art with Equal Error Rate (EER) of 2.98\% and tandem detection cost function (t-DCF) of 0.0817, surpassing most advanced anti-spoofing approaches.

\end{itemize}

The remainder of this paper is organized as follows. Section \uppercase\expandafter{\romannumeral2} introduces our speech anti-spoofing model. In Section \uppercase\expandafter{\romannumeral3} we present our experimental configuration, and our experimental results and discussions will be introduced in the Section \uppercase\expandafter{\romannumeral4}. Finally, we give our conclusion in Section \uppercase\expandafter{\romannumeral5}.

\begin{figure*}[t]
  \centering
  \includegraphics[width=\linewidth]{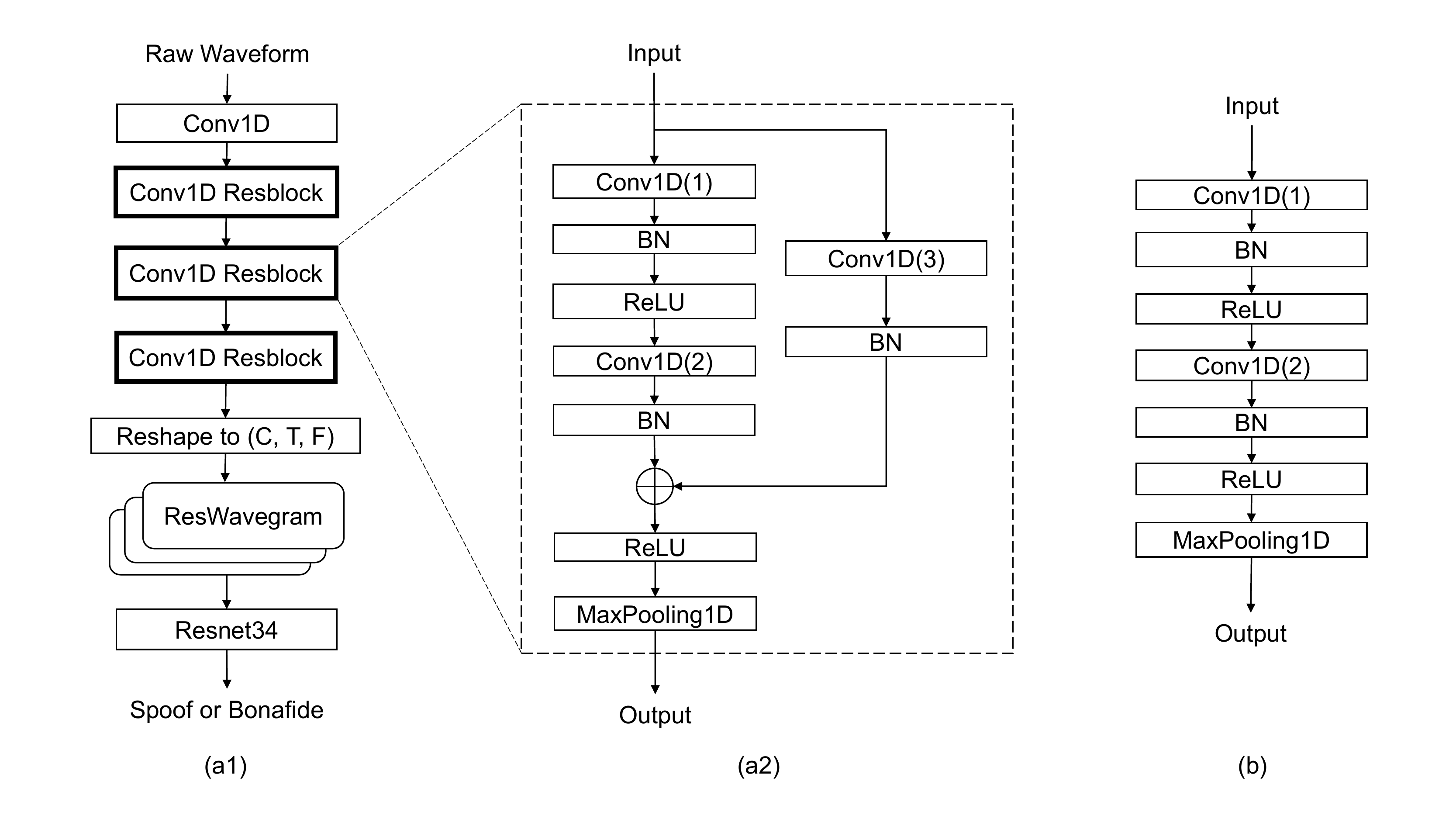}
  \caption{ResWavegram based anti-spoofing model. (a1) shows the ResWavegram feature used for the anti-spoofing model. (a2) presents our proposed Conv1D Resblock by adding the Conv1D(3) layer and the followed BN layer as a residual connection. The dilation of the Conv1D(1) and Conv1D(3) layer is set to 1, while that of the Conv1D(2) layer is set to 2. All kernel size of the convolution layers in the Conv1D Resblock is set to 3. (b) shows the details of the original Conv1D block.}
  \label{fig:Wavegram}
\end{figure*}

\section{Model Architecture}

Our anti-spoofing model mainly includes two parts: the feature extraction (Wavegram or ResWavegram) and the backbone (Resnet34). Before training, the time length of input audio is limited to 8 seconds by concatenating or cutting operations to ensure the same size of the input data. 
All audio samples are divided by 32768 ($2^{15}$), ensuring that their numerical range is restricted in [-1, 1]. Finally, the raw waveform is directly fed to the neural network and the whole model is trained end-to-end.

\subsection{Wavegram}
 Since the time-domain CNN models \cite{tak2020end, Jung2019} cannot capture the frequency relationships within an audio signal, it is important to construct a frequency axis from the one-dimensional CNNs based feature.
The extraction process of Wavegram is shown in Table~\ref{tab:wavegram}. The key structure to extract the Wavegram feature is the Conv1D block (Figure~\ref{fig:Wavegram} (b)). To improve the receptive field of the network, the Conv1D block consists of two convolution layers whose dilations are set to 1 and 2, respectively. Each Conv1D block contains a MaxPooling layer to down-sample the original audio signal.

The output size of an audio signal is changed from $1 \times 128000$ to $C \times T$ in a mini-batch. $C$ represents the final output channels, which equals $C_3$ in Table~\ref{tab:wavegram}. The number of the output time frame $T$ is $128000/(5\times4^3)=400$, while the stride of the Conv1D layer is set to 5 and the pooling size of the MaxPooling layer in each Conv1D block is set to 4. Then we reshape the feature size to $\frac{C}{F} \times T \times F$ by splitting the channels to $C_g=\frac{C}{F}$ group(s), where each group can be regarded as having $F$ frequency bins. In this way, each audio finally is transformed to a feature map whose size is of $C_g \times 400 \times F$. Then the feature is sent to the backbone Resnet34 to predict the input audio is either bonafide or spoofed.

%

\begin{table}[th]
  \caption{The extraction process of Wavegram. We can get different Wavegram based features by setting $(C_1, C_2, C_3)$ to different values. In our experiments, Wavegram-S, Wavegram-M and Wavegram-L features are got by setting $(C_1, C_2, C_3)$ to $(64, 64, 64)$, $(64, 128, 128)$ and $(64, 128, 256)$, respectively.}
  \label{tab:wavegram}
  \centering
    \begin{tabular}{c || c  }
    \hline
    Layer & Output Size \\
    \hline
    \hline
    Input & $1 \times 128000$ \\
    \hline
    Conv1D & $64 \times 25600$ \\
    \hline
    Conv1D block & $C_1 \times 6400$ \\
    \hline
    Conv1D block & $C_2 \times 1600$ \\
    \hline
    Conv1D block & $C_3 \times 400$ \\
    \hline
    Output(Wavegram) & $C_g \times 400 \times F$ \\
    \hline
\end{tabular}
\end{table}

\subsection{ResWavegram}

Inspired by the residual connection proposed in the Resnet network \cite{he2016deep}, we add a residual connection between the input and output of the two convolution layers in each Conv1D block. The modified block is shown in Figure~\ref{fig:Wavegram} (a2), and we call it Conv1D Resblock. The residual connection we add consists of a convolution layer and a batch normalization (BN) layer. In each Conv1D Resblock, the receiving field becomes larger with the kernel size of the Conv1D(3) layer is set to 3. A BN layer is added after the Conv1D (3) layer to speed up the training process and reduce the over-fitting effect of the network.

We get the ResWavegram feature by replacing the Conv1D block in the Wavegram extraction process with the Conv1D Resblock, as shown in Figure~\ref{fig:Wavegram} (a1). By adding such a residual connection, the perceptual information range of our feature extraction process is deepened and the information of gradient update can be better transmitted. The neural network ultimately learns a better feature representation from the raw waveform, which is proved by the experimental results.

\subsection{Resnet}
Resnet is widely used in sound-related tasks, and it is also a popular choice in the field of anti-spoofing. We believe that the Resnet can learn a more distinctive feature from the raw waveform. So, we use Resnet from \cite{he2016deep}, but the channels of each ResBlock are changed to 1/4 of the original ones. Two fully connected layers are added after the pooling layer, as shown in Table~\ref{tab:resnet34}. A skip connection is added between the outputs of the pooling layer and the FC2 layer to ensure that the gradient update information can be better passed through the network. We also add the Rectified Linear Unit (ReLU) activation function after the FC1 layer.

\begin{table}[th]
  \caption{Detailed architecture of Resnet. The Adaptive Average Pooling is selected as the pooling layer and the output size is set to (1,1), making each channel has one output fed to the fully connected layers.}
  \label{tab:resnet34}
  \centering
    \begin{tabular}{c || c | c | c}
    \hline
    Layer & Output Size & Channels & Blocks\\
    \hline
    \hline
    Conv\&BN & $T \times F$ &16 & -\\
    \hline
    Res1 & $T \times F$ &16 & 3\\
    \hline
    Res2 & $T/2 \times F/2$ &32 & 4\\
    \hline
    Res3 & $T/4 \times F/4$ &64 & 6\\
    \hline
    Res4 & $T/8 \times F/8$ &128 & 3\\
    \hline
    Pooling & 128 &- & -\\
    \hline
    FC1 & 128 &- & -\\
    \hline
    FC2 & 128 &- & -\\
    \hline
    Output & 2 &- & -\\
    \hline
\end{tabular}
\end{table}

\section{Experiments Setup}

\subsection{Database}
The database we use is the logical access (LA) corpus of ASVspoof2019 Challenge \cite{wang2020asvspoof}, which is based upon a multi-speaker speech synthesis database called VCTK \cite{yamagishi2019cstr}. All audio samples in the database are sampled at 16 kHz and 16-bit quantization. The database is divided into three subsets, namely train, development and evaluation, and the details are shown in Table~\ref{tab:database}. The evaluation subset contains a set of unseen bonafide and spoofed speech collected from multiple speakers, and the spoofed speech is generated by diverse unseen spoofing algorithms. We use the train and development subsets for joint training and the evaluation subset is used for evaluation, as many researchers do.

\begin{table}[th]
  \caption{Detailed information of the ASVspoof2019 LA corpus}
  \label{tab:database}
  \centering
    \begin{tabular}{ c || c | c | c | c }
    \hline
    Subset & Male & Female & Bonafide & Spoofed\\
    \hline
    \hline
    Train & 8 &12 & 2580 & 22800\\
    \hline
    Development & 4 &6 & 2548 & 22296\\
    \hline
    Evaluation & 21 &27 & 7355 & 63882\\
    \hline
\end{tabular}
\end{table}

\subsection{Model training}
We use Pytorch as the training platform and all the experiments are implemented with Python language.
We train each model with 50 epochs, while the batch size is set to 16. We select Adam \cite{kingma2015adam} as the optimizer with the initial learning rate of $1E{-4}$ and the weight decay of 0. The Cosine Annealing Warm Restarts algorithm \cite{loshchilov2016sgdr} is used for updating the learning rate and the learning rate is gradually down to $1E{-8}$ during the training process. Cross entropy (CE) loss is chosen as the loss function. Kaiming initialization \cite{he2015delving} is used for all the convolution layers while the weight and bias of the batch normalization (BN) layers are set to 1 and 0 respectively.

\subsection{Evaluation metrics}

We adopt the same two metrics that are used in ASVspoof2019 Challenge \cite{Todisco2019}. The primary metric used for evaluation is tandem detection cost function (t-DCF), ensuring the evaluation results will show the performance of spoofing countermeasures (CMs) on the reliability of an ASV system. Equal error rate (EER) is selected as the second evaluation metric, corresponding to the CM threshold where the miss and false alarm rates equal each other.

When doing evaluation, the score is calculated using the log-likelihood ratio. 
\begin{equation}
    score(U) = log(p(bonafide/U; \theta))-log(p(spoof/U; \theta)) 
\end{equation}
where $U$ represents the given audio signal to be tested and $\theta$ represents the parameters of our trained model.

\section{Results and Discussion}

\subsection{Baselines}
ASVspoof2019 Challenge introduces two baselines with the Gaussian mixture model (GMM) as the backend classifier. The two baselines use different acoustic frontends, namely linear frequency cepstral coefficients (LFCC) and constant Q cepstral coefficients (CQCC) \cite{Todisco2019}. The scores of these two official systems on the ASVspoof2019 LA evaluation set are shown in Table~\ref{tab:result_3}.

\subsection{Ablation studies}
To avoid the over-fitting and under-fitting situations during training, different channel combination configurations are explored to find a better combination for extracting the Wavegram. $(C_1, C_2, C_3)$ in Table~\ref{tab:wavegram} are set to $(64, 64, 64)$, $(64, 128, 128)$ and $(64, 128, 256)$ respectively. According to the combination configurations of the selected channels, we respectively named the three corresponding features as Wavegram-S, Wavegram-M and Wavegram-L. We select different $C_g$ in each model to explore its impact on the performance of our proposed anti-spoofing model. In our experiment, the values of $C_g$ are set to 1, 2, and 4, respectively. The ablation results are shown in Table~\ref{tab:result_1}.

In Table~\ref{tab:result_1}, we can see that setting $C_g=1$ usually achieves better results for the three models. One possible explanation is that all the extracted information can be placed in the dimension of the frequency when $C_g$ is set to 1, obtaining a more complete feature representation for the backbone Resnet34. The Wavegram-M based model always achieves the best results when select the same $C_g$ for these models, showing the channel combination of $(64, 128, 128)$ prevents our proposed model from the over-fitting and under-fitting situations. All scores got from these models are better than the official baseline systems.

\begin{table}[th]
  \caption{Wavegram based models tested on the ASVspoof2019 LA evaluation set. Wavegram-S, Wavegram-M and Wavegram-L with different values of $C_g$ are tested with the backbone Resnet34.}
  \label{tab:result_1}
  \centering
    \begin{tabular}{ c || c | c | c }
    \hline
    Feature & $C_g$ & t-DCF & EER(\%)\\
    \hline
    \hline
     & 1 & 0.1074 & 4.34 \\
    Wavegram-S & 2 &0.1240 & 5.52 \\
     & 4 &0.1121 & 4.56\\
    \hline
     & 1 & \textbf{0.0849} & \textbf{3.39} \\
    Wavegram-M & 2 &0.0922 & 3.44 \\
     & 4 &0.0938 & 3.68\\
    \hline
     & 1 &0.1007 & 4.01 \\
    Wavegram-L & 2 &0.0996 & 3.99 \\
     & 4 &0.1183 & 5.05\\
    \hline
\end{tabular}
\end{table}

\subsection{Comparative results}

Based on the experimental results on Wavegram, we set the value of $C_g$ to 1 to make better use of the learned raw waveform features. Comparative experiments are done with the value of $(C_1, C_2, C_3)$ changes. Like the extraction process of Wavegram, we can get the ResWavegram-S, ResWavegram-M and ResWavegram-L features by setting $(C_1, C_2, C_3)$ to $(64, 64, 64)$, $(64, 128, 128)$ and $(64, 128, 256)$, respectively. The final results of our experiments upon the Wavegram and our proposed ResWavegram are shown in Table~\ref{tab:result_2}.

\begin{table}[th]
  \caption{Comparative experiments of the Wavegram and ResWavegram tested on the ASVspoof2019 LA evaluation set. The value of $C_g$ is set to 1 in all the following models.}
  \label{tab:result_2}
  \centering
    \begin{tabular}{ c || c | c  }
    \hline
    Feature & t-DCF & EER(\%)\\
    \hline
    \hline
    Wavegram-S  & 0.1074 & 4.34 \\
    ResWavegram-S &0.0818 & 3.25 \\
    \hline
    Wavegram-M &  0.0849 & 3.39 \\ 
    ResWavegram-M & \textbf{0.0817} & \textbf{2.98} \\
    \hline
    Wavegram-L & 0.1007 & 4.01 \\
    ResWavegram-L & 0.0930 & 3.56 \\
    \hline
\end{tabular}
\end{table}

Compared to the Wavegram, ResWavegram achieves better performance under different configurations, especially for the ResWavegram-S feature. The network structure with the ResWavegram has much power to extract feature information from the waveform by using the residual connection in the Conv1d Resblock. Compared with the original Wavegram, the ResWavegram feature has approximately 12\% improvement on the EER evaluation metric. Our ResWavegram-Resnet (RW-Resnet) model improves the t-DCF and EER metrics of the best baseline LFCC-GMM by 61\% and 63\% respectively.

\begin{table}[th]
  \caption{Performance comparison of the proposed Wavegram-Resnet and RW-Resnet systems to some known single systems on the ASVspoof2019 LA evaluation set.}
  \label{tab:result_3}
  \centering
    \begin{tabular}{ c || c | c  }
    \hline
    System & t-DCF & EER(\%)\\
    \hline
    \hline
    CQCC-GMM (Baseline1) \cite{Todisco2019}& 0.237 &9.57\\
    LFCC-GMM (Baseline2) \cite{Todisco2019}& 0.212 &8.09 \\
    \hline
    MFCC-Resnet \cite{Alzantot2019}& 0.204 & 9.33\\
    Spec-Resnet \cite{Alzantot2019}& 0.274 & 9.68\\
    CQCC-Resnet \cite{Alzantot2019}& 0.217 & 7.69\\
    LFCC-Siamese CNN \cite{Lei2020} & 0.093 &3.79\\
    Optimised LFCC (full-band) \cite{Tak2020} & 0.090 & 3.50\\
    LFCC-LCNN \cite{Lavrentyeva2019}& 0.100 & 5.06\\
    FFT-LCNN \cite{Lavrentyeva2019}& 0.103 & 4.53\\
    FG-LCNN \cite{Wu2020}& 0.102 & 4.07\\
    S1-RawNet2 \cite{tak2020end}& 0.130 & 5.64\\
    S2-RawNet2 \cite{tak2020end}& 0.118 & 5.13\\
    S3-RawNet2 \cite{tak2020end}& 0.129 & 4.66\\
    \hline
    Wavegram-Resnet (Ours) &  0.085 & 3.39 \\ 
    RW-Resnet (Ours) & \textbf{0.082} & \textbf{2.98} \\
    \hline
\end{tabular}
\end{table}

As shown in Table~\ref{tab:result_3}, the proposed Wavegram-Resnet and RW-Resnet models for synthetic speech attacks detection are compared with other advanced systems tested on the ASVspoof2019 LA evaluation set \cite{wang2020asvspoof}. Compared with neural networks that use traditional methods to extract features \cite{Lavrentyeva2019, Alzantot2019, Wu2020}, the performances of our proposed models are greatly improved. Compared with the RawNet2 that only uses one-dimensional convolution \cite{tak2020end}, our method of using two-dimensional convolution has better feature performance capabilities. Compared with the well-designed system with the Siamese CNN \cite{Lei2020}, our proposed models achieve better performance using a slightly modified Resnet34 as the backbone. The performances of the two models we proposed far exceed the two baselines \cite{Todisco2019} and other recently proposed single systems on the ASVspoof2019 LA evaluation set. Finally, we compare the score of our proposed RW-Resnet single system with the top5 fusion systems in ASVspoof2019 Challenge, as shown in Table~\ref{tab:result_4}.

\begin{table}[th]
  \caption{Performance comparison of our proposed RW-Resnet single system to the top5 fusion systems in ASVspoof2019 Challenge on the ASVspoof2019 LA evaluation set.}
  \label{tab:result_4}
  \centering
    \begin{tabular}{ c || c | c  }
    \hline
    System & t-DCF & EER(\%)\\
    \hline
    \hline
    T05 & 0.0069 & 0.22\\
    T45 \cite{Lavrentyeva2019} & 0.0510 & 1.86\\
    T60 \cite{Chettri2019} & 0.0755 & 2.64\\
    RW-Resnet (Ours) & \textbf{0.0817} & \textbf{2.98} \\
    T24 & 0.0953 & 3.45\\
    T50 \cite{Yang2019} & 0.1118 & 3.56\\
    \hline
\end{tabular}
\end{table}

\section{Conclusions}
A novel speech anti-spoofing model named RW-Resnet is introduced in this paper. We propose the Conv1D Resblock with a residual connection, which allows the model to learn a better feature representation from the raw waveform. The effectiveness of directly feeding the raw waveform into a neural network, with no information of the raw waveform is discard, has been verified by comparing with other advanced speech anti-spoofing systems. Our RW-Resnet improves the t-DCF and EER metrics of the best baseline LFCC-GMM by 61\% and 63\% respectively. In the future, we intend to adapt our proposed method to other sound fields, and explore more expressive raw-waveform based models.


\bibliographystyle{IEEEtran}

\bibliography{main}

\end{document}